\documentclass[runningheads,a4paper]{llncs}

\usepackage{amssymb}
\usepackage{graphicx}

\usepackage{url}
\urldef{\mailsa}\path|{alfred.hofmann, ursula.barth, ingrid.haas, frank.holzwarth,|
\urldef{\mailsb}\path|anna.kramer, leonie.kunz, christine.reiss, nicole.sator,|
\urldef{\mailsc}\path|erika.siebert-cole, peter.strasser, lncs}@springer.com|    
\newcommand{\keywords}[1]{\par\addvspace\baselineskip
\noindent\keywordname\enspace\ignorespaces#1}

\begin{document}

\mainmatter  

\title{PFMFind: a system for discovery of peptide homology and function}

\titlerunning{PFMFind: a system for discovery of peptide homology and function}

%
%
\author{Aleksandar Stojmirovi\'c\inst{1}\thanks{to whom correspondence should be addressed} %
   \and Peter Andreae\inst{2} %
   \and Mike Boland\inst{3} %
   \and Thomas William Jordan\inst{4} %
   \and Vladimir G. Pestov\inst{5}}

\authorrunning{Stojmirovi\'c \emph{et al.}}


\institute{National Center for Biotechnology Information, National Library of Medicine, National Institutes of Health, Bethesda, MD 20894, United States %
\and  School of Engineering and Computer Science, Victoria University of Wellington, PO Box 600, Wellington 6140, New Zealand %
\and  Riddet Institute, Massey University, PB 11 222, Palmerston North 4442, New Zealand %
\and  School of Biological Sciences, Victoria University of Wellington, PO Box 600, Wellington 6140, New Zealand %
\and  Department of Mathematics and Statistics, University of Ottawa, 585 King Edward Ave., Ottawa, ON K1N 6N5, Canada}

%
%
%
%

%
%

\toctitle{PFMFind: a system for discovery of peptide homology and function}
\tocauthor{Aleksandar Stojmirovi\'c, Peter Andreae, Mike Boland, Thomas William Jordan, Vladimir G. Pestov}
\maketitle

\begin{abstract}
Protein Fragment Motif Finder (PFMFind) is a system that enables efficient
discovery of relationships between short fragments of protein sequences using
similarity search. It supports queries based on amino acid similarity matrices  
and position specific score matrices (PSSMs) obtained through an iterative procedure. 
PSSM construction is customisable through plugins written in Python. PFMFind
consists of a GUI client, an index for fast similarity
search and a relational database for storing search results and sequence
annotations. It is written mostly in Python. The components of PFMFind 
communicate through TCP/IP sockets and can be located on different physical
machines. PFMFind is freely available for download (under a GPL licence) 
from \\
\url{http://pfmfind.stojmirovic.org}

\end{abstract}

\section{Introduction}

The biological functions of proteins are as much a function of particular motifs of peptide sequence as they are of the overall protein structure.  It is of interest to the biologist to search for examples of convergent motifs as they are likely to indicate a functional role. While many approaches exist for finding longer sequence motifs (50 amino acids or more), finding relationships between short fragments (3--18 amino acids long) of full protein sequences also promises great rewards
in understanding novel aspects of protein structure and function. These
relationships might be evolutionary in origin or might arise by
convergence, that is, by acquisition of the same biological function in
evolutionarily distant species.  

Finding short motifs presents significant challenges because many of the apparent relationships between short fragments could have arisen by chance and thus have no functional significance. Furthermore, most widely available tools for sequence database search and motif finding were designed with longer motifs in mind. For example, Watt and Doyle~\cite{WD05} observed that the NCBI BLAST~\cite{AMSZ97} family of programs, the best known set of tools for searching biological sequence datasets, is not suitable for
identifying shorter sequences with particular constraints and proposed
a pattern search tool to find DNA or protein fragments that match a
given sequence or a pattern exactly. This paper outlines
the Protein Fragment Motif Finder (PMFind), a tool that uses database search to
identify the conserved short peptide motifs of a query sequence and associates them with the available functional annotations. 

\section{Overview}

\begin{figure}[!tpb]
\centerline{\scalebox{0.9}{\includegraphics{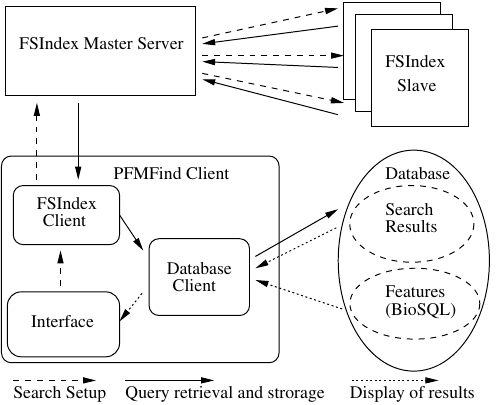}}}
\caption{Structure of PFMFind system}\label{fig:PFMFind_struct}
\end{figure}

The PFMFind system consists of three major components: a search
engine for fast similarity search of datasets of short peptide
fragments called FSIndex, a relational database, and the PFMFind graphical user interface
(GUI) client (Fig.~\ref{fig:PFMFind_struct}). 
PFMFind client takes user input, and communicates with FSIndex and the database 
through its components. It passes search parameters in batches to FSIndex and 
receives the results of searches that are then stored in the database.
It also retrieves the results from the database and displays them,
together with available annotations, to the user. The annotations are
stored in a separate (BioSQL) schema in the database.

Most of PFMFind was written in the Python programming language,
and uses both the standard Python library and additional
modules such as Biopython
(\url{http://www.biopython.org}). The components
communicate using the standard TCP/IP socket interface and can therefore be
located on different machines. Since PFMFind is highly modular, the GUI client
can be replaced by a Python script for non-interactive use.

\subsection{Similarity Search}

PFMFind supports searches of datasets of short peptide fragments of
fixed length using an ungapped similarity score obtained by summing
similarity scores at each position of the fragments being compared.
The positional similarity scores can be defined by standard score
matrices such as PAM~\cite{DSO78} or BLOSUM~\cite{HH92}, or 
by position specific score matrices (PSSMs)~\cite{GME87}. A fragment dataset 
consists of all fragments of a specified length from a given protein sequence 
dataset (where the fragments may overlap).

Iterative construction of PSSM, similar to that used by PSI-BLAST~\cite{AMSZ97}, 
is supported through plugins -- Python routines
that take the results of a previous search and construct a PSSM. The default
plugin uses the weighting procedure of Henikoff and Henikoff
\cite{HH94} to assign weights to fragments and Dirichlet mixtures
\cite{SKBH96} for regularising the amino acid frequency counts at each
position. Users with some knowledge of Python can create their own plugins and
use them for searches by placing then in the appropriate directory.

Search criteria can be specified according to threshold raw similarity
scores, distances, p-values and E-values (range search), as well as the number of
closest data points to retrieve (kNN search). The probability model for calculation 
of p-values assumes that the score of each fragment is the sum of
independent random variables corresponding to the score at each
position and the score distribution is calculated using discrete Fast Fourier Transform.

\subsection{FSIndex}

The heart of PFMFind is FSIndex, an efficient indexing scheme for
similarity search of very large datasets of short protein fragments of fixed
length~\cite{PS06,SP07}. FSIndex is based on two principles: reduction of the amino
acid alphabet to clusters largely based on their biochemical
properties (hydrophobic, polar, charged, aromatic ...) and
combinatorial generation of neighbours.  The design of FSIndex means
that a typical search involves scanning less than 1\% of the fragment
dataset, but ensures that no neighbours satisfying search criteria are
ever missed.

FSIndex is implemented in the C programming language and embedded into
Python, with the whole data structure as well as the indexed sequences
stored in primary memory. For even greater efficiency, computation of
searches can be distributed among several machines using a
master/slave model: the master handles p-value computations,
distributes queries to slaves, each of which is indexing a different
part of the dataset, and communicates with the client. 

\subsection{Database}

The second major component of PFMFind is a relational database, used
both for storage of search results and the sequence annotation.  We
use PostgreSQL, a freely available modern database management system.

Each user of the system has their own schema for storing search
results.  The database also stores all search parameters, including
PSSMs and the results of each iteration, facilitating reversion to a
previous iteration without repeating the whole procedure.

The database stores sequence annotations in a standard BioSQL schema
available to all users.  PFMFind also contains scripts for loading
four types of information beyond the protein sequence:
Uniprot~\cite{BAWB05} keywords and features, Uniref clusters~\cite{BAWB05} and InterPro~\cite{HJMA11} domains. 
When retrieved for display, annotations are joined to search results through accession numbers. 

\subsection{GUI Client}

The final PFMFind component is a GUI client that connects to both the
FSIndex master and the database component. To perform fragment
searches, the user specifies a query sequence, usually a long sequence
that is broken into overlapping fragments of fixed length, and chooses
the fragment lengths, threshold parameters and the actual fragments in
the query sequences that will be used for the search. 

\begin{figure}[!tpb]
\centerline{\scalebox{0.27}{\includegraphics{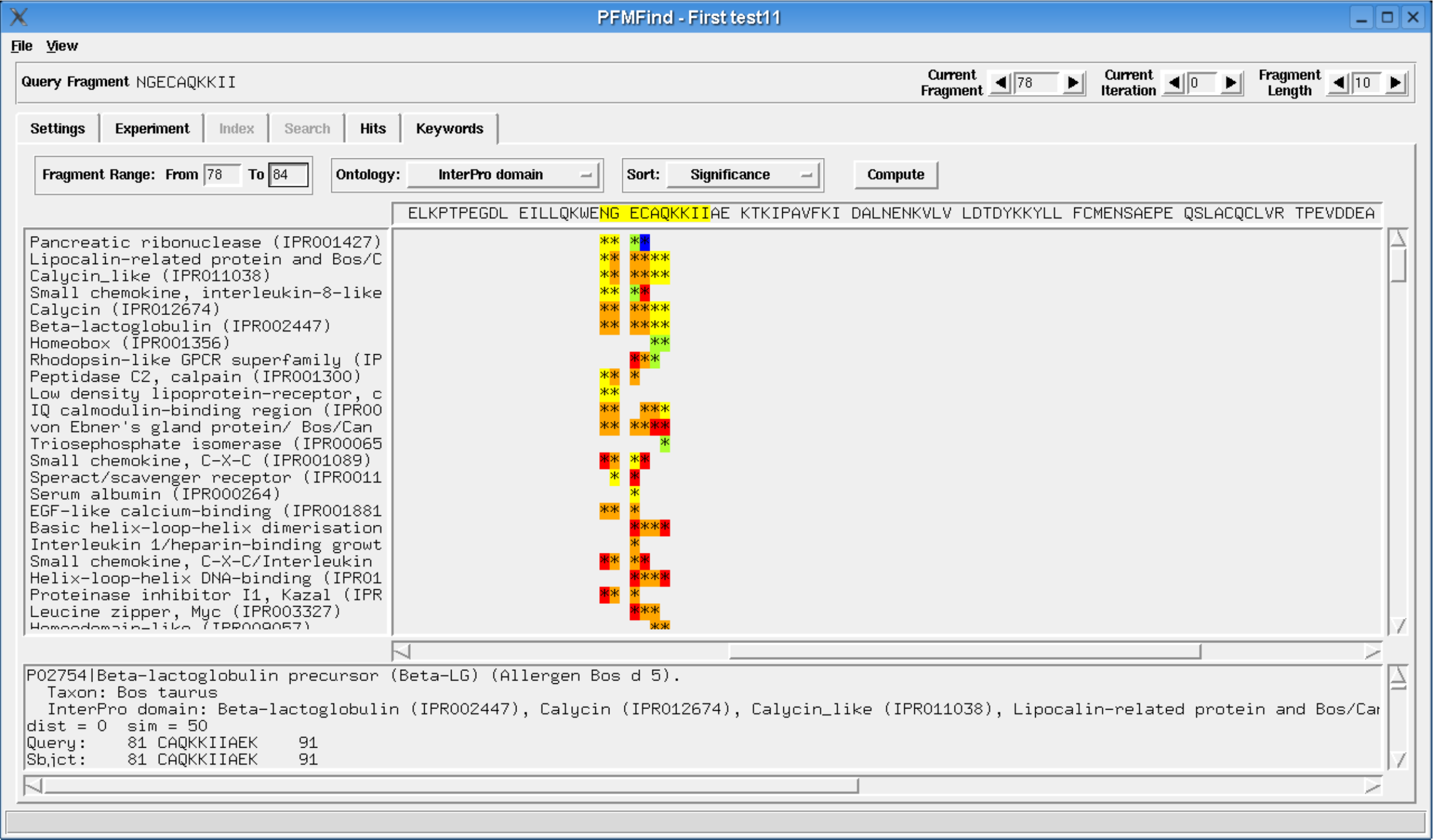}}}
\caption{A screenshot of PFMFind GUI showing search results associated with their annotations}\label{fig:snaphshot}
\end{figure}

The GUI client can display search results both as lists of hits associated with a
particular location in the query sequence and as a feature vs location
dot plot -- each location matching a particular feature is marked by
a coloured dot (Fig.~\ref{fig:snaphshot}). Dots are colour coded by the number of hits matching
the feature to distinguish frequently represented features from those
that appear only a few times in the hit list. The GUI client also performs
all computations for constructing PSSMs.

\section{Conclusion}

PFMFind is an efficient, flexible, and extensible framework for similarity
search of datasets of short peptide fragments. It supports fast similarity
search with selectivity and sensitivity specified by
PSSMs and associates search results with biological function by
using sequence features and annotations.

\section*{Acknowledgements}
We wish to thank Pavle Mogin and Danyl McLauchlan for their help with PostgreSQL and testing the software, respectively. A.S. was supported by a Bright Future PhD scholarship awarded by the NZ
Tertiary Education Commission jointly with the Fonterra Research Centre, 
by a Fields Institute/University of Ottawa postdoctoral fellowship, and by the Intramural Research Program of the National Library of Medicine at the National Institutes of Health. V.G.P. and A.S. acknowledge
support from NSERC discovery grant program and University of Ottawa internal grants.

\bibliographystyle{splncs03}
\bibliography{pfmfind}

\begin{thebibliography}{10}
\providecommand{\url}[1]{\texttt{#1}}
\providecommand{\urlprefix}{URL }

\bibitem{AMSZ97}
Altschul, S.F., Madden, T.L., Schaffer, A.A., Zhang, J., Zhang, Z., Miller, W.,
  Lipman, D.J.: Gapped {BLAST} and {PSI--{BLAST}}: a new generation of protein
  database search programs. Nucleic Acids Res  25,  3389--3402 (1997)

\bibitem{BAWB05}
Bairoch, A., Apweiler, R., Wu, C.H., Barker, W.C., Boeckmann, B., Ferro, S.,
  Gasteiger, E., Huang, H., Lopez, R., Magrane, M., Martin, M.J., Natale, D.A.,
  O'Donovan, C., Redaschi, N., Yeh, L.S.L.: {The Universal Protein Resource
  (UniProt)}. Nucleic Acids Res  33 Database Issue,  154--159 (2005)

\bibitem{DSO78}
Dayhoff, M.O., Schwartz, R.M., Orcutt, B.C.: A model of evolutionary change in
  proteins. In: Dayhoff, M.O. (ed.) Atlas of Protein Sequence and Structure,
  vol.~5, chap.~22, pp. 345--352. National Biomedical Research Foundation
  (1978)

\bibitem{GME87}
Gribskov, M., McLachlan, A.D., Eisenberg, D.: Profile analysis: detection of
  distantly related proteins. Proc Natl Acad Sci USA  84,  4355--4358 (1987)

\bibitem{HH94}
Henikoff, S., Henikoff, J.G.: {Position-based sequence weights}. J Mol Biol
  243(4),  574--578 (1994)

\bibitem{HH92}
Henikoff, S., Henikoff, J.: Amino acid substitution matrices from protein
  blocks. Proc Natl Acad Sci USA  89,  10915--10919 (1992)

\bibitem{HJMA11}
Hunter, S., Jones, P., Mitchell, A., Apweiler, R., Attwood, T.K., Bateman, A.,
  Bernard, T., Binns, D., Bork, P., Burge, S., de~Castro, E., Coggill, P.,
  Corbett, M., Das, U., Daugherty, L., Duquenne, L., Finn, R.D., Fraser, M.,
  Gough, J., Haft, D., Hulo, N., Kahn, D., Kelly, E., Letunic, I., Lonsdale,
  D., Lopez, R., Madera, M., Maslen, J., McAnulla, C., McDowall, J., McMenamin,
  C., Mi, H., Mutowo-Muellenet, P., Mulder, N., Natale, D., Orengo, C.,
  Pesseat, S., Punta, M., Quinn, A.F., Rivoire, C., Sangrador-Vegas, A.,
  Selengut, J.D., Sigrist, C.J.A., Scheremetjew, M., Tate, J.,
  Thimmajanarthanan, M., Thomas, P.D., Wu, C.H., Yeats, C., Yong, S.Y.:
  Interpro in 2011: new developments in the family and domain prediction
  database. Nucleic Acids Res  40(Database issue),  D306--12 (Jan 2012)

\bibitem{PS06}
Pestov, V., Stojmirovi\'c, A.: Indexing schemes for similarity search: an
  illustrated paradigm. Fundam. Inform.  70(4),  367--385 (2006)

\bibitem{SKBH96}
Sjölander, K., Karplus, K., Brown, M., Hughey, R., Krogh, A., Mian, I.,
  Haussler, D.: Dirichlet mixtures: A method for improving detection of weak
  but significant protein sequence homology. Comput. Appl. Biosci.  12(4),
  327--345 (1996)

\bibitem{SP07}
Stojmirovi\'c, A., Pestov, V.: Indexing schemes for similarity search in
  datasets of short protein fragments. Inf. Syst.  32(8),  1145--1165 (2007)

\bibitem{WD05}
Watt, T.J., Doyle, D.F.: {ESPSearch: a program for finding exact sequences and
  patterns in DNA, RNA, or protein}. Biotechniques  38(1),  109--115 (2005)

\end{thebibliography}

\end{document}